\documentclass[conference]{IEEEtran}
\IEEEoverridecommandlockouts
\usepackage{cite}
\usepackage{amsmath,amssymb,amsfonts,amsthm}
\usepackage{mathtools}

\newtheorem{definition}{Definition}

\usepackage{algorithmic}
\usepackage{graphicx}
\usepackage{textcomp}
\usepackage{xcolor}
\def\BibTeX{{\rm B\kern-.05em{\sc i\kern-.025em b}\kern-.08em
    T\kern-.1667em\lower.7ex\hbox{E}\kern-.125emX}}

\usepackage[hidelinks]{hyperref}
\usepackage{hyperref}

\usepackage{booktabs}
\usepackage{multirow}

\usepackage{blindtext}

\usepackage[acronym,ucmark,translate]{glossaries}
\setacronymstyle{long-short}
\glsdisablehyper

\newacronym{ai}{AI}{Artificial Intelligence}
\newacronym{iot}{IoT}{Internet-of-Things}
\newacronym{asic}{ASIC}{Application-Specific Integrated Circuit}
\newacronym{cnn}{CNN}{Convolutional Neural Network}
\newacronym{conv}{CONV}{Convolution}
\newacronym{nn}{NN}{Neural Network}
\newacronym{dnn}{DNN}{Deep Neural Network}
\newacronym{rnn}{RNN}{Recurrent Neural Network}
\newacronym{mac}{MAC}{Multiply-Accumulate}
\newacronym{nas}{NAS}{Neural Architecture Search}
\newacronym{soc}{SoC}{System-on-Chip}
\newacronym{fpga}{FPGA}{Field-Programmable Gate Array}
\newacronym{phy}{PHY}{Physical layer}
\newacronym{imu}{IMU}{Inertial Measurement Unit}
\newacronym{ble}{BLE}{Bluetooth Low Energy}
\newacronym{lstm}{LSTM}{Long Short-Term Memory}
\newacronym{ctc}{CTC}{Connectionist Temporal Classification}
\newacronym{isa}{ISA}{Instruction Set Architecture}
\newacronym{cer}{CER}{Character Error Rate}
\newacronym{gatt}{GATT}{General ATTribute profile}
\newacronym{mtu}{MTU}{Maximum Transmission Unit}
\newacronym{ci}{CI}{Connection Interval}
\newacronym{ll}{LL}{Link Layer}
\newacronym{lm}{LM}{Language Model}
\newacronym{fpu}{FPU}{Floating-Point Unit}
\newacronym{tcn}{TCN}{Temporal Convolutional Network}
\newacronym{ucie}{UCIe}{Universal Chiplet Interconnect Express}
\newacronym{eda}{EDA}{Electronic Design Automation}
\newacronym{onnx}{ONNX}{Open Neural Network Exchange}
\newacronym{ga}{GA}{Genetic Algorithm}
\newacronym{dla}{DLA}{Deep Learning Accelerator}
\newacronym{cots}{COTS}{Commercial Off-The-Shelf}
\newacronym{qat}{QAT}{Quantization-Aware Training}
\newacronym{dag}{DAG}{Directed Acyclic Graph}

\usepackage{tikz}
\usepackage{pgfplots}
\pgfplotsset{compat=1.18}
\usepackage{pgfplotstable}
\usepackage{subcaption}
\usepackage{fontawesome5}
\usepackage{relsize}

\usepackage[american,siunitx, RPvoltages]{circuitikz}

\usetikzlibrary{arrows, positioning, shapes.geometric, chains, arrows.meta}
\usetikzlibrary{decorations.pathreplacing, matrix, fit}
\usetikzlibrary{decorations.pathreplacing,angles,quotes}
\usetikzlibrary{automata}

\usetikzlibrary{calc,shapes,shadows,patterns,backgrounds}
\usepgfplotslibrary{groupplots, units}

\tikzset{
    every picture/.style={
        /utils/exec={\sffamily}
    },
    box/.style={
        rectangle, fill=KITblue_30, inner sep=4pt, minimum width=#1, align=center, minimum height=6.0mm,
        font=\footnotesize, rounded corners=1pt, draw=KITblack_70,
        execute at begin node=\setlength{\baselineskip}{10pt}
    },
    plaintext/.style={
        box, fill=none, draw=none
    },
    envbox/.style={
        box, draw=KITblack_50, dashed, fill=gray!10!white, line width=2pt
    },
    line/.style={
        thick, rounded corners=1pt
    },
    arrow/.style={
        line, -stealth
    }
}
\usepackage[]{xcolor}

\definecolor{KITgreen_100}{RGB}{0,150,130}
\definecolor{KITgreen_70}{RGB}{76,181,167}
\definecolor{KITgreen_50}{RGB}{127,202,192}
\definecolor{KITgreen_30}{RGB}{178,223,217}
\definecolor{KITgreen_15}{RGB}{217,239,236}

\definecolor{KITblue_100}{RGB}{70,100,170}
\definecolor{KITblue_70}{RGB}{125,146,195}
\definecolor{KITblue_50}{RGB}{162,177,212}
\definecolor{KITblue_30}{RGB}{199,208,229}
\definecolor{KITblue_15}{RGB}{227,232,242}

\definecolor{KITmaygreen_100}{RGB}{140,182,60}
\definecolor{KITmaygreen_70}{RGB}{171,204,108}
\definecolor{KITmaygreen_50}{RGB}{193,219,147}
\definecolor{KITmaygreen_30}{RGB}{216,232,186}
\definecolor{KITmaygreen_15}{RGB}{235,245,220}

\definecolor{KITyellow_100}{RGB}{252,229,0}
\definecolor{KITyellow_70}{RGB}{252,237,76}
\definecolor{KITyellow_50}{RGB}{252,240,126}
\definecolor{KITyellow_30}{RGB}{255,247,179}
\definecolor{KITyellow_15}{RGB}{255,252,217}

\definecolor{KITorange_100}{RGB}{223,155,27}
\definecolor{KITorange_70}{RGB}{232,183,88}
\definecolor{KITorange_50}{RGB}{237,202,133}
\definecolor{KITorange_30}{RGB}{245,223,181}
\definecolor{KITorange_15}{RGB}{250,237,217}

\definecolor{KITbrown_100}{RGB}{167,130,45}
\definecolor{KITbrown_70}{RGB}{191,163,94}
\definecolor{KITbrown_50}{RGB}{209,186,132}
\definecolor{KITbrown_30}{RGB}{227,211,177}
\definecolor{KITbrown_15}{RGB}{242,235,216}

\definecolor{KITred_100}{RGB}{162,34,35}
\definecolor{KITred_70}{RGB}{191,86,86}
\definecolor{KITred_50}{RGB}{209,125,128}
\definecolor{KITred_30}{RGB}{227,172,175}
\definecolor{KITred_15}{RGB}{242,213,213}

\definecolor{KITpurple_100}{RGB}{163,16,124}
\definecolor{KITpurple_70}{RGB}{191,69,159}
\definecolor{KITpurple_50}{RGB}{209,113,184}
\definecolor{KITpurple_30}{RGB}{227,166,211}
\definecolor{KITpurple_15}{RGB}{242,208,233}

\definecolor{KITzyanblue_100}{RGB}{35,161,224}
\definecolor{KITzyanblue_70}{RGB}{96,188,235}
\definecolor{KITzyanblue_50}{RGB}{139,204,240}
\definecolor{KITzyanblue_30}{RGB}{184,225,245}
\definecolor{KITzyanblue_15}{RGB}{217,240,250}

\definecolor{KITblack_100}{RGB}{0,0,0}
\definecolor{KITblack_70}{RGB}{77,77,77}
\definecolor{KITblack_50}{RGB}{128,128,128}
\definecolor{KITblack_30}{RGB}{179,179,179}
\definecolor{KITblack_15}{RGB}{217,217,217}

\newcommand\copyrighttext{%
  \footnotesize \textcopyright 2024 IEEE. Personal use of this material is permitted.
  Permission from IEEE must be obtained for all other uses, in any current or future
  media, including reprinting/republishing this material for advertising or promotional
  purposes, creating new collective works, for resale or redistribution to servers or
  lists, or reuse of any copyrighted component of this work in other works.
  DOI: \href{https://doi.org/10.1109/ISVLSI61997.2024.00019}{10.1109/ISVLSI61997.2024.00019}}
\newcommand\copyrightnotice{%
\begin{tikzpicture}[remember picture,overlay]
\node[anchor=south,yshift=10pt] at (current page.south) {\fbox{\parbox{\dimexpr\textwidth-\fboxsep-\fboxrule\relax}{\copyrighttext}}};
\end{tikzpicture}%
}

\begin{document}

\title{Automated Deep Neural Network Inference Partitioning for Distributed Embedded Systems}

\author{\IEEEauthorblockN{
Fabian Kre\ss,
El Mahdi El Annabi,
Tim Hotfilter,
Julian Hoefer,
Tanja Harbaum,
Juergen Becker}
\IEEEauthorblockA{Karlsruhe Institute of Technology, Karlsruhe, Germany \\
\{fabian.kress, hotfilter, julian.hoefer, harbaum, becker\}@kit.edu}
}

\maketitle
\copyrightnotice

\begin{abstract}
Distributed systems can be found in various applications, e.g., in robotics or autonomous driving, to achieve higher flexibility and robustness. Thereby, data flow centric applications such as Deep Neural Network (DNN) inference benefit from partitioning the workload over multiple compute nodes in terms of performance and energy-efficiency. However, mapping large models on distributed embedded systems is a complex task, due to low latency and high throughput requirements combined with strict energy and memory constraints.

In this paper, we present a novel approach for hardware-aware layer scheduling of DNN inference in distributed embedded systems. Therefore, our proposed framework uses a graph-based algorithm to automatically find beneficial partitioning points in a given DNN. Each of these is evaluated based on several essential system metrics such as accuracy and memory utilization, while considering the respective system constraints.
We demonstrate our approach in terms of the impact of inference partitioning on various performance metrics of six different DNNs. As an example, we can achieve a 47.5\% throughput increase for EfficientNet-B0 inference partitioned onto two platforms while observing high energy-efficiency.
\end{abstract}


\section{Introduction}
The hardware requirements for \gls{dnn}-based applications such as Advanced Driver-Assistance Systems (ADAS) or assistance robots are constantly increasing with regard to various performance indicators including latency, throughput and energy efficiency.
To address this, several dedicated hardware accelerator architectures have been proposed over the last years.
However, these are either tailored to a specific workload offering high performance but less flexibility, or designed to accelerate various applications but suffer from low efficiency.
As a result, systems consisting of more than a single hardware accelerator are required to offer well-fitted hardware for different applications executed on embedded platforms.
Distributed systems in those use cases consist of several hardware platforms close to the sensors and a central compute unit.
Implementing multiple hardware accelerators for \gls{dnn} inference in a single component thereby leads to a significant increase in design complexity to avoid severe bottlenecks introduced through limited memory bandwidth.
As a result, distributing hardware accelerators for \gls{dnn} inference is advantageous \cite{DeepThings}.

However, mapping \glspl{dnn} on such systems in a performant and energy-efficient manner is a complex task, as the architecture of the hardware accelerators must be taken into account in addition to the transmission overhead.
In this paper, we address this problem by proposing an automated inference partitioning approach to find an optimal trade-off considering several performance metrics and also hardware constraints.
Our proposed framework thereby uses a graph-representation of the \gls{dnn} to find a Pareto-optimal mapping, which offers low latency, high energy-efficiency and high \gls{dnn} accuracy.
It supports multiple hardware accelerators tailored to different workloads and also considers the overhead introduced by the link between the system platforms.
In summary, our contributions are as follows:

\begin{itemize}
    \item We present our automated framework to evaluate several important metrics of potential \gls{dnn} partitioning points.
    \item We introduce our systematic approach to determine an optimal partitioning based on given constraints and the optimization goals.
    \item We evaluate our methodology for several \glspl{cnn} and present our experimental results regarding latency, energy consumption, throughput, top-1 accuracy and memory consumption.
\end{itemize}

\section{Related Work}
In recent years, partitioning of \gls{dnn} inference has been studied in different contexts.
Several research has been done on inference partitioning of multi-FPGA environments~\cite{Alonso21},~\cite{Agut23}.
They focus on efficient use of shared memory resources between FPGAs to reduce communication overhead and to increase performance. 
Others \cite{DeepThings, Kakolyris23} perform partitioning in edge clusters and thereby try to distribute the \gls{dnn} inference in parallel over multiple devices. 

\begin{table*}[t]
    \renewcommand{\arraystretch}{1.2}
    \centering
    \scalebox{0.85}{
    \begin{tabular}{r|cccccccl}
        \multirow{2}{*}{\textbf{Framework}} & \textbf{DNN} & \textbf{Search} & \textbf{Auto.} & \textbf{Multi} & \textbf{HW-aware} & \multirow{2}{*}{\textbf{(Re-)Training}} & \textbf{Target} & \multirow{2}{*}{\textbf{Optimization Metrics}}  \\
        & \textbf{Framework} & \textbf{Method} & \textbf{Part.} & \textbf{Part.} & \textbf{Part.} & & \textbf{Platform} & \\
        \toprule
        
        Hu et al. \cite{Hu19} & Caffe & Graph-based & \faCheck & \faTimes & \faTimes & \faTimes & RPi 3 & latency, throughput \\

        \midrule
        Neurosurgeon \cite{Neurosurgeon} & Caffe & Model-based & \faCheck & \faTimes & \faTimes & \faTimes & CPU/GPU & latency, energy \\

        \midrule
        Yao et al. \cite{Yao20} & TensorFlow & Model-based & \faCheck & \faTimes & \faTimes & \faCheck & CPU/GPU & latency, bandwidth \\



        \midrule
        Ko et al. \cite{Ko18} & \textit{N/A} & Simulation & \faTimes & \faTimes & \faCheck & \faCheck & ASIC & energy, throughput, bandwidth, accuracy \\
        \midrule
        CNNParted \cite{cnnparted} & PyTorch & Simulation & \faTimes & \faTimes & \faCheck & \faTimes & ASIC/GPU & latency, energy, bandwidth \\ 
        
        \midrule
        AxoNN \cite{axonn} & TensorFlow & Graph-based & \faCheck & \faCheck & \faCheck & \faTimes & GPU/DLA & latency, energy \\ 
        
        \midrule
        Our Proposal & ONNX & Graph-based & \faCheck & \faCheck & \faCheck & \faCheck & ASIC & latency, bandwidth, energy, memory, accuracy, throughput \\       
        \bottomrule
    \end{tabular}}
    \vspace{0.25cm}
    \caption{State-of-the-art methodologies for evaluation of \gls{dnn} inference partitioning in the edge} 
    \vspace{-0.45cm}
    \label{tab:soa}
\end{table*}

These approaches are not suitable for systems that do not consist of a computing cluster to distribute the workload.
Consequently, there has also been research on layer-wise \gls{dnn} inference partitioning as shown in \autoref{tab:soa}.
Hu et al.~\cite{Hu19} proposed an automated dynamic scheduling approach which takes network conditions into account. 
The graph-based algorithm thereby either tries to optimize the inference towards minimum latency or maximum throughput, depending on the network load.
In the evaluation, the authors can show the benefits of partitioning \gls{dnn} inference, however, their scheduling algorithm does not take hardware performance and limitations into account and is therefore not suited for resource constrained systems.
Moreover, the scheduler Neurosurgeon~\cite{Neurosurgeon} has been proposed to optimize \gls{dnn} inference partitioning between mobile edge and cloud. 
It is designed to automatically determine a beneficial partitioning point in terms of overall latency and mobile edge energy consumption.
Similarly to the previously mentioned approach, their scheduling does not provide a hardware-optimized \gls{dnn} inference partitioning.

Apart from optimizing latency, energy consumption and throughput by applying inference partitioning, reducing bandwidth is also a major objective in this context.
Therefore, Yao et al. \cite{Yao20} proposed an offloading approach which adds an autoencoder structure at the partitioning point to further compress the data transmitted between computing nodes in the system.
Since this approach can impact the overall accuracy of the \gls{dnn}, training of the autoencoder has to be done with regard to all potential partitioning points.
Again, the methodology does not take hardware constraints into account and is only evaluated on two mobile devices.
In contrast to previous methodologies, Ko et al. \cite{Ko18} proposed a simulation-based \gls{dnn} inference partitioning approach for more resource-constrained and \gls{asic} platforms. 
The presented methodology thereby allows to determine a partitioning point based on energy efficiency and throughput.
To further reduce the bandwidth required when distributing inference, the authors propose to encode intermediate feature maps using lossy encoding and fine-tuning of the partitioned \gls{dnn}.
However, the methodology presented only allows to derive design guidelines for other applications and does not include an algorithm for automated partitioning.
Similarly, the open-source tool CNNParted \cite{cnnparted} offers a hardware-aware design space exploration of beneficial partitioning points but only outputs latency, energy and bandwidth metrics of each potential point to the designer.
Nevertheless, since it is based on open-source tools for modeling and evaluating arbitrary \gls{dnn} hardware accelerators, the exploration also allows to compare different architectures during design time.

However, none of these approaches considers multiple partitioning points during inference.
AxoNN \cite{axonn} aims to find near-optimal schedules for multi-accelerator based \glspl{soc} on \gls{cots} platforms such as Nvidia Jetson Xavier AGX containing a GPU and a \gls{dla}.
In particular, the transition costs between the available hardware accelerators in a shared-memory system are taken into account during evaluation by the authors, highlighting that near-optimal schedules mostly include only a single transition between GPU and \gls{dla}.
The evaluation of partitioning, however, only takes latency and energy consumption into account. As a result, this approach is not suitable for distributed embedded systems.

\section{Problem Statement}
The scheduling of \gls{dnn} inference depends primarily on the system architecture, i.e. the number of available hardware accelerators in the system. 
In the simplest case, we only have two computing platforms available for executing \gls{dnn} inference and can define a partitioning point as follows: 
\begin{definition}[Partitioning Point] \label{def:partpoint}
    In a system consisting of two hardware platforms $A$ and $B$, a partitioning point is a layer $l_p$ with $p \in \{1, \dots, L \}$ of a given \gls{dnn} with $L$ layers, such that:
    \begin{itemize}
        \item Each layer $l_i$ with $i \in \{1, \dots, L\}$ is executed once per inference on exactly one platform in the system.
        \item Each layer $l_i$ with $i \in \{0, \dots, p\}$ is executed on $A$.
        \item Each layer $l_i$ with $i \in \{p + 1, \dots, L\}$ is executed on $B$.
        \item The intermediate feature map $f_p$ of $l_p$ is transmitted over a link from $A$ to $B$.
    \end{itemize}
\end{definition}
Determining an optimal partitioning point under given constraints depends on the main optimization goal which is application specific.
As a result, we formulate the objective functions as a minimization problem based on weighted sum of cost functions:
\begin{definition}[Minimization Problem] \label{def:prob}
    The minimization problem for a system consisting of two hardware platforms $A$ and $B$ is given as
    \begin{equation*}
        \underset{l_p=0\dots L}{\operatorname*{minimize}} \sum_{i=0}^{N} c_i\cdot \theta_i(l_p)
    \end{equation*}
    with L being the number of \gls{dnn} layers, $N$ the number of optimization criteria, and  $c_i$ being application dependent coefficients of the cost functions $\theta_i$.
\end{definition}

Based on the state of the art, we find that existing frameworks cover overall five different optimization metrics for determining an optimal partitioning point: latency $d(l_p)$, throughput $th(l_p)$, energy consumption $e(l_p)$, link bandwidth $bw(l_p)$, and accuracy $acc(l_p)$. 
With the emergence of large-scale models and the further increasing complexity of \glspl{dnn}~\cite{Desislavov23}, memory size is becoming a potential bottleneck in embedded systems as well.
Therefore, solving the optimization problem for real-world hardware requires to include the required amount of memory $m_A(l_p)$ and $m_B(l_p)$ for storing \gls{dnn} parameters and intermediate feature maps on platform $A$ and $B$, respectively. 
Each of the different metrics is modeled as a cost function $\theta_0 \dots \theta_N$ depending on the partitioning point and can be constrained as part of the minimization problem.

\section{Our Proposed Framework}
Even for a system configuration that features only two hardware accelerators, determining an optimal partitioning point is a non-trivial task.
In this section, we therefore present our framework for exploring the design space of potential partitioning points that meet the previously stated objectives.
An overview of our approach is given in \autoref{fig:framework}.
As input, the framework takes a \gls{dnn} description as \gls{onnx} file, problem constraints, and the main optimization objective.
Using \gls{onnx} as input allows to combine our proposed framework with a wide range of \gls{dnn} frameworks such as TensorFlow or PyTorch.

\definecolor{analysis_color}{RGB}{169,221,240}
\definecolor{analysis_box_color}{RGB}{200,220,170}

\definecolor{filter_color}{RGB}{210,239,219}

\definecolor{quant_color}{RGB}{204,193,219}
\definecolor{quant_box_color}{RGB}{169,199,221} 

\definecolor{hweval_color}{RGB}{237,202,133}
\definecolor{hweval_box_color}{RGB}{255,182,173}

\definecolor{scheduling_color}{RGB}{0,145,156}
\definecolor{scheduling_box_color}{RGB}{200,222,182} 

\begin{figure}[t]
    \centering
    \scalebox{0.66}{
    \begin{tikzpicture}
    [
        scale=1.0, every node/.style={scale=1.0}
    ]
    \renewcommand{\arraystretch}{0.8}

    \node[box, fill=analysis_color, minimum width=5cm, minimum height=1.5cm] at (0,0) (analysis_box) {};
    \node[above, inner sep=6pt, text=black, font=\bfseries, rotate=90] at (analysis_box.east) (analysis_box_txt) {\scriptsize{Analysis}};
    \node[box, fill=analysis_box_color, minimum width=1.75cm, minimum height=1cm] at ([xshift=0.85cm]analysis_box.center) (graphana) {\footnotesize{Graph}\\\footnotesize{Analysis}};
    \node[box, fill=analysis_box_color, minimum width=1.75cm, minimum height=1cm] at ([xshift=-1.25cm] analysis_box.center) (graphinit) {\footnotesize{Graph}\\\footnotesize{Initialization}};
    \node[plaintext] at ([xshift=-1cm]analysis_box.west) (onnx_file) {\huge{\faIcon[regular]{file-code}}};
    \node[plaintext] at (onnx_file.north) (onnx_file_txt) {\scriptsize{ONNX}};

    \draw[arrow] (onnx_file.east) -- (graphinit.west);
    \draw[arrow] (graphinit.east) -- (graphana.west);

    \node[box, fill=filter_color, minimum width=5cm, minimum height=1.5cm] at ([xshift=0mm, yshift=-2.3cm] analysis_box.south) (filter_box) {};
    \node[above, inner sep=6pt, text=black, font=\bfseries, rotate=90] at (filter_box.east) (filter_box_txt) {\scriptsize{Filter}};
    \node[box, fill=white, minimum width=1.75cm, minimum height=1cm] at ([xshift=0.85cm]filter_box.center) (memeval) {\footnotesize{Memory}\\\footnotesize{Evaluation}};
    \node[plaintext] at ([xshift=0.65cm,yshift=0.85cm]memeval.north) (mem_const_file) {\huge{\faIcon[regular]{file-alt}}};
    \node[plaintext] at ([yshift=0.15cm]mem_const_file.north) (mem_const_file_txt) {\scriptsize{Memory}\\[-1mm]\scriptsize{Constraints}};
    \node[box, fill=white, minimum width=1.75cm, minimum height=1cm] at ([xshift=-1.25cm] filter_box.center) (linkeval) {\footnotesize{Link}\\\footnotesize{Evaluation}};
    \node[plaintext] at ([xshift=-0.65cm,yshift=0.85cm]linkeval.north) (link_const_file) {\huge{\faIcon[regular]{file-alt}}};
    \node[plaintext] at ([yshift=0.15cm]link_const_file.north) (link_const_file_txt) {\scriptsize{Link}\\[-1mm]\scriptsize{Constraints}};
    \node[plaintext] at ([xshift=-1cm]filter_box.west) (linkmodel_file) {\huge{\faIcon[regular]{file-code}}};
    \node[plaintext] at ([yshift=0.15cm]linkmodel_file.north) (linkmodel_file_txt) {\scriptsize{Link}\\[-1mm]\scriptsize{Model}};

    \draw[arrow] (graphana.south) -- (memeval.north);
    \draw[arrow] (link_const_file.south) -- ([xshift=-0.65cm]linkeval.north);
    \draw[arrow] (mem_const_file.south) -- ([xshift=0.65cm]memeval.north);
    \draw[arrow] (memeval.west) -- (linkeval.east);
    \draw[arrow] (linkmodel_file.east) -- (linkeval.west);

    \node[box, fill=quant_color, minimum width=6cm, minimum height=4.5cm] at ([xshift=-5mm, yshift=-4cm] filter_box.south) (quant_box) {};
    \node[above, inner sep=6pt, text=black, font=\bfseries, rotate=90] at (quant_box.east) (quant_box_txt) {\scriptsize{Quantization}};
    \node[box, fill=quant_box_color, minimum width=1.75cm, minimum height=1cm] at ([xshift=-1.25cm, yshift=1.5cm] quant_box.center) (calib) {\footnotesize{Parameter}\\\footnotesize{Calibration}};
    \node[box, fill=quant_box_color, minimum width=1.75cm, minimum height=1cm] at ([xshift=-1.25cm, yshift=0cm] quant_box.center) (quant) {\footnotesize{Quantization}};
    \node[box, fill=quant_box_color, minimum width=1.75cm, minimum height=1cm] at ([xshift=-1.25cm, yshift=-1.5cm] quant_box.center) (retrain) {\footnotesize{Retraining}};
    \node[box, fill=white, minimum width=1.75cm, minimum height=1cm] at ([xshift=1.35cm, yshift=-0.75cm] quant_box.center) (acceval) {\footnotesize{Accuracy}\\\footnotesize{Evaluation}};
    \node[plaintext] at ([xshift=0.4cm,yshift=3.1cm]acceval.north) (acc_const_file) {\huge{\faIcon[regular]{file-alt}}};
    \node[plaintext] at ([yshift=0.15cm]acc_const_file.north) (acc_const_file_txt) {\scriptsize{Accuracy}\\[-1mm]\scriptsize{Constraint}};

    \draw[arrow] (quant.south) -- (retrain.north);
    \draw[arrow] (quant.south) -- (retrain.north);
    \draw[arrow] ([yshift=0.25cm]acceval.west) -- +(-0.425,0) |- (quant.east);
    \draw[arrow] (retrain.east) -| +(0.425,0.25) |- ([yshift=-0.25cm]acceval.west);
    
    \draw[arrow] (calib.east) -| ([xshift=-0.4cm]acceval.north);
    \draw[arrow] ([xshift=0.5cm]linkeval.south) |- +(0,-1.5) -| (acceval.north);
    \draw[arrow] (acc_const_file.south) -- ([xshift=0.4cm]acceval.north);

    \draw[line] (graphinit.south) -- ([yshift=0.25cm]linkeval.north);
    \draw[arrow] ([yshift=-0.25cm]linkeval.south) |- +(-0.1,-1.5) -| (calib.north);

    \node[box, fill=hweval_color, minimum width=5cm, minimum height=1.5cm] at ([xshift=5mm, yshift=-2.5cm] quant_box.south) (hweval_box) {};
    \node[above, inner sep=6pt, text=black, font=\bfseries, rotate=90] at (hweval_box.east) (hweval_box_txt) {\scriptsize{HW Eval}};
    \node[box, fill=white, minimum width=1.75cm, minimum height=1cm] at ([xshift=0.85cm]hweval_box.center) (perfeval) {\footnotesize{Performance}\\\footnotesize{Evaluation}};
    \node[plaintext] at ([yshift=0.85cm]perfeval.north) (hw_const_file) {\huge{\faIcon[regular]{file-alt}}};
    \node[plaintext] at ([yshift=0.15cm]hw_const_file.north) (hw_const_file_txt) {\scriptsize{Latency, Throughput}\\[-1mm]\scriptsize{Energy Constraints}};
    \node[box, fill=hweval_box_color, minimum width=1.75cm, minimum height=1cm] at ([xshift=-1.25cm] hweval_box.center) (probmap) {\footnotesize{Problem}\\\footnotesize{Mapping}};
    \node[plaintext] at ([xshift=-1cm]hweval_box.west) (hwmodel_file) {\huge{\faIcon[regular]{file-code}}};
    \node[plaintext] at ([yshift=0.15cm]hwmodel_file.north) (hwmodel_file_txt) {\scriptsize{HW}\\[-1mm]\scriptsize{Models}};

    \draw[arrow] (acceval.south) |- +(0,-1.2) -| (probmap.north);
    \draw[arrow] (hw_const_file.south) -- (perfeval.north);
    \draw[arrow] (hwmodel_file.east) -- (probmap.west);
    \draw[arrow] (probmap.east) -- (perfeval.west);

    \node[box, fill=scheduling_color, minimum width=6cm, minimum height=1.5cm] at ([xshift=-5mm,yshift=-1.3cm]hweval_box.south) (sched_box) {};
    \node[above, inner sep=6pt, text=white, font=\bfseries, rotate=90] at (sched_box.east) (sched_txt) {\scriptsize{Scheduling}};
    \node[box, fill=scheduling_box_color, minimum width=4.5cm, minimum height=1cm] at (sched_box.center) (eval) {\footnotesize{Evaluating Pareto-optimal Points}};
    \node[plaintext] at ([xshift=-1cm]sched_box.west) (opt_file) {\huge{\faIcon[regular]{file-code}}};
    \node[plaintext] at ([yshift=0.15cm]opt_file.north) (opt_file_txt) {\scriptsize{Optimization}\\[-1mm]\scriptsize{Goal}};

    \draw[arrow] ([xshift=-0.5cm]linkeval.south) |- +(-2.5,-1.5) |- +(-2,-10) -| ([xshift=-1.5cm]eval.north);
    \draw[arrow] (acceval.south) |- +(1.8,-1.2) |- +(0,-4.5) -| ([xshift=0.5cm]eval.north);
    \draw[arrow] (perfeval.south) -- +(0,-0.4) -| ([xshift=-0.5cm]eval.north);
    \draw[arrow] (memeval.south) |- +(2,-0.4) |- +(1.8,-10.1) -| ([xshift=1.5cm]eval.north);
    \draw[arrow] (opt_file.east) -- (eval.west);

    \node[plaintext] at  ([xshift=-1cm, yshift=-0.8cm]sched_box.south) (output) {\huge{\faIcon{file-alt}}};
    \node[plaintext] at ([yshift=-0.15cm]output.south) (output_txt) {\scriptsize{Partitioning}\\[-1mm]\scriptsize{Point}};
    \node[plaintext] at  ([xshift=1cm, yshift=-0.8cm]sched_box.south) (output2) {\huge{\faIcon{chart-bar}}};
    \node[plaintext] at ([yshift=-0.22cm]output2.south) (output2_txt) {\scriptsize{Partitioning}\\[-1mm]\scriptsize{Statistics}};
    
    \draw[arrow] ([xshift=-1cm]eval.south) -- (output.north);
    \draw[arrow] ([xshift=1cm]eval.south) -- (output2.north);

    \end{tikzpicture}
    }
    \caption{Overview of our proposed framework. First, a graph is generated based on the \gls{onnx} description. After filtering of potential partitioning points considering memory and link constraints, quantization is performed and evaluated. Finally, the framework estimates performance on hardware and selects a Pareto-optimal point.}
    \label{fig:framework}
\end{figure}
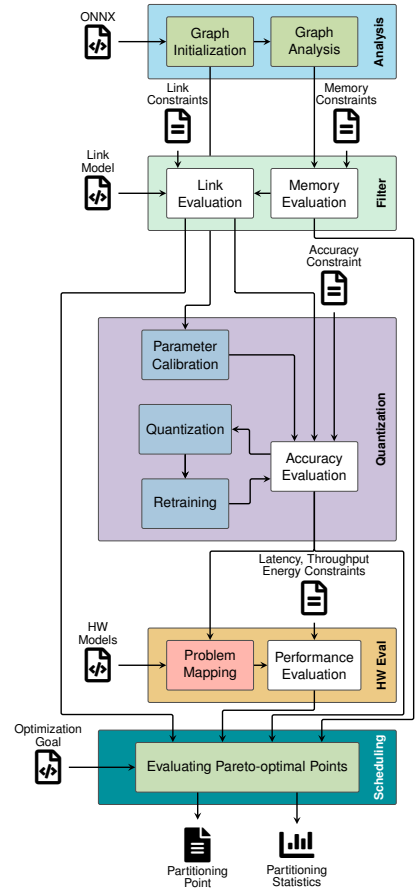

First, the \gls{dnn} is converted into a graph representation and then analyzed to find all potential partitioning points without considering given constraints.
For reducing the number of feasible schedules of the \gls{dnn}, these partitioning points are filtered based on memory and link evaluation. 
This step thereby requires a link model to determine bottlenecks in the link that would otherwise violate system constraints.
Furthermore, since hardware accelerators often use integer or fixed-point representation instead of floating-point values for calculation, quantization has to be applied for \gls{dnn} inference in hardware.
However, accuracy of the \gls{dnn} can be significantly impacted by using such number formats.
Consequently, the accuracy drop is evaluated for each remaining partitioning point, optionally including retraining.

The list of all partitioning points not violating accuracy, bandwidth or memory constraints is then forwarded to the Hardware (HW) Evaluation step, where latency, throughput and energy consumption metrics are evaluated for different schedules.
Thereby, the performance of hardware accelerators depends on the mapping for each layer and the used process technology.
We use Timeloop \cite{timeloop} and Accelergy \cite{accelergy} to find a near-optimal mapping for each layer to the hardware accelerator and to estimate latency and energy consumption. 

In the next step, the collected metrics are used to evaluate all potential partitioning points fulfilling the given constraints.
Since we perform multi-objective optimization, we use the NSGA-II to determine Pareto-optimal points in the resulting set of potential partitioning points ~\cite{pymoo}.
Thereby, the partitioning point serves as variable of the partitioning problem.
Since the complexity of a \gls{dnn} varies significantly, the population size as well as the number of generations is set depending on the number of layers.
Finally, the framework identifies the most favorable \gls{dnn} partitioning point that fulfills the main optimization objective.

\subsection{Graph Analysis}
Finding potential partitioning points in a given \gls{dnn} can be done in different ways as shown in \autoref{tab:soa}.
Simulation-based approaches such as CNNParted \cite{cnnparted} try to determine branches in the network architecture based on comparing output and input shapes of layers.
However, since this approach is only analyzing the \gls{dnn} structure on basic layer level, simple mathematical operations such as adding, concatenating and flattening tensors have to be identified using complex logic. 
In contrast, our proposed framework is able to directly extract the graph representation from the \gls{onnx} specification.

Furthermore, skip connections are often used to address the vanishing gradient problem.
As a result, the \gls{dag} of a given \gls{dnn} can consist of parallel branches that can be executed independently.
Our framework therefore first performs a topological sort of the \gls{dag} to find a linear ordering of its vertices. 
This step is required to determine a schedule for executing the \gls{dnn} inference sequentially on each of the available hardware accelerators.
In case there are parallel branches, the algorithm randomly selects one of the unscheduled layers as the next node to be added to the linear sequence.


\subsection{Memory Size Estimation}
In many recent \glspl{soc}, memory allocates a significant part of die area.
At the same time, the complexity of \glspl{dnn} is constantly rising and, thus, the memory demand.
As a result, we also have to consider this when looking for a feasible schedule for \gls{dnn} inference.
The required memory size for a sequence of layers without branches is formalized as follows:
\begin{definition}[Memory Size] \label{def:memsize}
    The memory size required to execute non-pipelined branch-free \gls{dnn} inference from layer $l_n$ to $l_m$ on a hardware accelerator $A$ is given by
    \begin{equation*}
        m_A(l_n, l_m) = \left( \sum_{i=n}^{m} s_i + \operatorname*{max}(a_n, \ldots, a_m ) \right) \cdot b_A, \quad n \leq m
    \end{equation*}
    \begin{equation*}
        a_{j} = f_{j,in} + f_{j,out} 
    \end{equation*}
    with $s_i$ being the number of parameters of layer $l_i$, $a_j$ the sum of input and output feature map size $f_{j,in}$ and $f_{j,out}$ of layer~$j$, and $b_A$ the quantized bit width of parameters and feature maps.
\end{definition}

Based on this, the memory requirements of each individual layer can be calculated.
As soon as the resulting demand exceeds the available on-chip memory of a platform, all following potential partitioning points are removed from the list of feasible schedulings.
However, in case of branches in the \gls{dnn} topology, different schedules have to be evaluated since parallel layers can be executed in different orders.
Therefore, the framework builds subgraphs for these parallel branches to find the schedule with minimum memory requirements according to \autoref{def:memsize}.

\subsection{Accuracy Exploration}


Hardware floating-point operations are not as energy efficient as integer computations.
As a result, hardware accelerators in embedded systems typically use integer or fixed-point \gls{mac} operations~\cite{simba, eyeriss}.
While \gls{dnn} models generally tolerate quantization very well, radical quantization schemes worsen the model accuracy.
Hence, quantization also has to be considered for finding a beneficial inference schedule in multi-accelerator systems. 

Before the actual exploration, our tool has to perform a parameter calibration to determine the ranges of features maps and weights.
In the next step, the impact of quantization on the model accuracy for each potential partitioning point has to be evaluated.
To determine the resulting accuracy, our proposed framework uses fake quantization, which allows to obtain the results quickly.
The degree of quantization thereby depends on the hardware specification of the implemented accelerator in each part of the system. 
However, \gls{dnn} accuracy can suffer significantly from quantization, especially in cases where a small bit width is used for inference in the hardware accelerator.
As a result, our framework offers the possibility to run \gls{qat} based on the trained network parameters to restore the accuracy of the model.

\subsection{Throughput Estimation}
Applications such as autonomous driving have certain throughput requirements to enable reliable decision making.
Running \gls{dnn} inference on a single accelerator might provide suitable latency but low throughput. 
Therefore, it must be considered as well when searching for an optimal partitioning point of a \gls{dnn}.
Since we assume that the hardware platforms can work in parallel as an asynchronous pipeline, the throughput is determined by the platform with the highest latency.
As a result, it can be formalized as follows:

\begin{definition}[Throughput] \label{def:throughput}
    The throughput for partitioned \gls{dnn} inference on a system consisting of two hardware platforms $A$ and $B$ connected over a $link$ is given by
    \begin{equation*}
        th(l_p) = min \left( \frac{1}{d_A}, \frac{1}{d_{Link}}, \frac{1}{d_B} \right)
    \end{equation*}
    with $d$ being the latency of each involved module of the system.
\end{definition}
\section{Evaluation}

Finally, we show experimental results for inference partitioning using our proposed framework.
The explorations have been conducted on a system consisting of a 64-core AMD EPYC 9554 and an NVIDIA RTX A6000 running Rocky Linux.
To evaluate the latency and energy consumption of each hardware accelerator, Timeloop is configured to use a linear-pruned search algorithm and a victory condition of 100.

\subsection{Workload}
First, we performed several explorations for a system model consisting of two hardware platforms A and B according to the problem as described in \autoref{def:prob}.
Platform A is based on a 16-bit Eyeriss-like architecture \cite{eyeriss} running at 200~MHz (EYR), and platform B uses a Simba-like accelerator~\cite{simba} at 200~MHz (SMB).
In our simulated system model, the platforms are connected to each other via Gigabit Ethernet.
To explore the transition costs of the link, i.e. latency and energy consumption, we used the open-source model provided by CNNParted \cite{cnnparted}.
We tested our framework for six feed-forward image classification \glspl{cnn}, i.e. EfficientNet-B0, ResNet-50, RegNetX\_400MF, VGG-16, GoogLeNet, and SqueezeNet V1.1 pretrained on ImageNet dataset.
In the quantization stage, we performed two epochs of \gls{qat} with 1,281,167~images for training and validated the quantized models with 8,192 images from the ImageNet dataset.

\begin{figure*}[t]

\newcommand\figwidth{6.0}
\newcommand\figheight{3.95}
\newcommand\figrelwidth{0.31}
\centering
    \begin{subfigure}[t]{\figrelwidth\linewidth}
    \centering
    \caption{\footnotesize{VGG-16}}
    \begin{tikzpicture}
        \begin{axis}[
            xlabel={\scriptsize{Latency [ms]}},
            ylabel={\scriptsize{Energy [mJ]}},
            width=\figwidth cm,
            height=\figheight cm,
            scatter/classes={%
                dom={mark=x,scale=0.85,draw=black!60},
                nondom={mark=oplus,scale=0.85,draw=red, fill=red},
                sensor={mark=square*, scale=1,draw=orange ,fill=orange},
                edge={mark=square*, scale=1,draw=blue,fill=blue},
                opt={mark=triangle*, scale=2,draw=teal,fill=teal},
                opt2={mark=triangle*, scale=2,draw=olive,fill=olive}
            },
            x label style={at={(axis description cs:0.5,-0.1)}, anchor=north},
            y label style={at={(axis description cs:-.25,0.5)}, anchor=north},
            tick label style={font=\scriptsize},
            yticklabel style={
                /pgf/number format/fixed,
                /pgf/number format/precision=1
            },
            scaled y ticks=false
        ]
        
        \addplot[scatter, only marks, scatter src=explicit symbolic] table[col sep=comma,header=false,x index=5,y index=6, meta index=11]{figures/eval_vgg16_result.tex};
        \node[coordinate,pin={[pin distance=0.2cm]-90:{\scriptsize\color{orange}EYR}}] at (axis cs:766.4029579999999,118.52051243756249) {};
        \node[coordinate,pin={[pin distance=0.2cm]90:{\scriptsize\color{blue}SMB}}] at (axis cs:302.75364599999995,84.30655697985384) {};
        \node[coordinate,pin={[pin distance=0.2cm]0:{\scriptsize\color{olive}ReLu\_1}}] at (axis cs:300.62592,83.11694982273431) {};
        \node[coordinate,pin={[pin distance=0.2cm]0:{\scriptsize\color{teal}ReLu\_2}}] at (axis cs:360.83712,77.07222337636603) {};
    
        \end{axis}
    \end{tikzpicture}
    \end{subfigure}\hfill
    \begin{subfigure}[t]{\figrelwidth\linewidth}
    \centering
    \caption{\footnotesize{ResNet-50}}
    \begin{tikzpicture}
        \begin{axis}[
            xlabel={\scriptsize{Energy [mJ]}},
            ylabel={\scriptsize{Throughput [FPS]}},
            width=\figwidth cm,
            height=\figheight cm,
            scatter/classes={%
                nondom={mark=oplus,scale=0.85,draw=red, fill=red},
                dom={mark=x,scale=0.85,draw=black!60},
                sensor={mark=square*, scale=1,draw=orange ,fill=orange},
                edge={mark=square*, scale=1,draw=blue,fill=blue},
                opt={mark=triangle*, scale=2,draw=teal,fill=teal},
                opt2={mark=triangle*, scale=2,draw=olive,fill=olive}
            },
            x label style={at={(axis description cs:0.5,-0.1)}, anchor=north},
            y label style={at={(axis description cs:-.25,0.5)}, anchor=north},
            tick label style={font=\scriptsize},
            yticklabel style={
                /pgf/number format/fixed,
                /pgf/number format/precision=2
            },
            scaled y ticks=false
        ]
        
        \addplot[scatter, only marks, scatter src=explicit symbolic] table[col sep=comma,header=false,x index=6,y index=7, meta index=11]{figures/eval_resnet50_result.tex};
        \node[coordinate,pin={[pin distance=0.2cm]180:{\scriptsize\color{orange}EYR}}] at (axis cs:35.825554064983024,4.833989208989211) {};
        \node[coordinate,pin={[pin distance=0.2cm]-90:{\scriptsize\color{blue}SMB}}] at (axis cs:23.635469507690402,12.407686810132617) {};
        \node[coordinate,pin={[pin distance=0.2cm]0:{\scriptsize\color{teal}ReLu\_11}}] at (axis cs:37.09306311116511,16.007909187771492) {};
    
        \end{axis}
    \end{tikzpicture}
    \end{subfigure}\hfill
    \begin{subfigure}[t]{\figrelwidth\linewidth}
    \centering
    \caption{\footnotesize{ResNet-50}}
    \begin{tikzpicture}
        \begin{axis}[
            xlabel={\scriptsize{Top-1 Accuracy [\%]}},
            ylabel={\scriptsize{Throughput[FPS]}},
            width=\figwidth cm,
            height=\figheight cm,
            scatter/classes={%
                dom={mark=x,scale=0.85,draw=black!60},
                nondom={mark=oplus,scale=0.85,draw=red, fill=red},
                sensor={mark=square*, scale=1,draw=orange ,fill=orange},
                edge={mark=square*, scale=1,draw=blue,fill=blue},
                opt={mark=triangle*, scale=2,draw=teal,fill=teal},
                opt2={mark=triangle*, scale=2,draw=olive,fill=olive}
            },
            x label style={at={(axis description cs:0.5,-0.1)}, anchor=north},
            y label style={at={(axis description cs:-.25,0.5)}, anchor=north},
            tick label style={font=\scriptsize},
            yticklabel style={
                /pgf/number format/fixed,
                /pgf/number format/precision=1
            },
             x filter/.code={\pgfmathparse{#1*100}\pgfmathresult},
            scaled y ticks=false
        ]
        
        \addplot[scatter, only marks, scatter src=explicit symbolic] table[col sep=comma,header=false,x index=11,y index=7, meta index=12]{figures/eval_resnet50_result_acc.tex};
        \node[coordinate,pin={[pin distance=0.25cm]90:{\scriptsize\color{orange}EYR}}] at (axis cs:87.34,4.833989208989211) {};
        \node[coordinate,pin={[pin distance=0.15cm]-90:{\scriptsize\color{blue}SMB}}] at (axis cs:85.50,12.407686810132617) {};
        \node[coordinate,pin={[pin distance=0.15cm]0:{\scriptsize\color{teal}Add\_3}}] at (axis cs:86.72,15.509608512665764) {};

        \end{axis}
    \end{tikzpicture}
    \end{subfigure}\hfill
    \begin{subfigure}[t]{\figrelwidth\linewidth}
    \centering
    \caption{\footnotesize{SqueezeNet V1.1}}
    \begin{tikzpicture}
        \begin{axis}[
            xlabel={\scriptsize{Latency [ms]}},
            ylabel={\scriptsize{Energy [mJ]}},
            width=\figwidth cm,
            height=\figheight cm,
            scatter/classes={%
                nondom={mark=oplus,scale=0.85,draw=red, fill=red},
                dom={mark=x,scale=0.85,draw=black!60},
                sensor={mark=square*, scale=1,draw=orange ,fill=orange},
                edge={mark=square*, scale=1,draw=blue,fill=blue},
                opt={mark=triangle*, scale=2,draw=teal,fill=teal},
                opt2={mark=triangle*, scale=2,draw=olive,fill=olive}
            },
            x label style={at={(axis description cs:0.5,-0.1)}, anchor=north},
            y label style={at={(axis description cs:-.25,0.5)}, anchor=north},
            tick label style={font=\scriptsize},
            yticklabel style={
                /pgf/number format/fixed,
                /pgf/number format/precision=1
            },
            scaled y ticks=false,
        ]
        
        \addplot[scatter, only marks, scatter src=explicit symbolic] table[col sep=comma,header=false,x index=5,y index=6, meta index=11]{figures/eval_squeezenet1_1_result.tex};
        \node[coordinate,pin={[pin distance=0.15cm]-90:{\scriptsize\color{orange}EYR}}] at (axis cs:16.955048000000005,3.4592531977525196) {};
        \node[coordinate,pin={[pin distance=0.1cm]0:{\scriptsize\color{blue}SMB}}] at (axis cs:10.233322999999999,3.703410938204118) {};
        \node[coordinate,pin={[pin distance=0.4cm]90:{\scriptsize\color{teal}ReLu\_2}}] at (axis cs:9.296391999999999,2.7765819353158716) {};
    
        \end{axis}
    \end{tikzpicture}
    \end{subfigure}\hfill
    \begin{subfigure}[t]{\figrelwidth\linewidth}
    \centering
    \caption{\footnotesize{EfficientNet-B0}}
    \begin{tikzpicture}
        \begin{axis}[
            xlabel={\scriptsize{Energy [mJ]}},
            ylabel={\scriptsize{Throughput [FPS]}},
            width=\figwidth cm,
            height=\figheight cm,
            scatter/classes={%
                dom={mark=x,scale=0.85,draw=black!60},
                nondom={mark=oplus,scale=0.85,draw=red, fill=red},
                sensor={mark=square*, scale=1,draw=orange ,fill=orange},
                edge={mark=square*, scale=1,draw=blue,fill=blue},
                opt={mark=triangle*, scale=2,draw=teal,fill=teal},
                opt2={mark=triangle*, scale=2,draw=olive,fill=olive}
            },
            legend style={
                font=\scriptsize, 
                at={(1, -0.32)},
                legend columns=2,
                column sep=0.3em},
            x label style={at={(axis description cs:0.5,-0.1)}, anchor=north},
            y label style={at={(axis description cs:-.25,0.5)}, anchor=north},
            tick label style={font=\scriptsize},
            yticklabel style={
                /pgf/number format/fixed,
                /pgf/number format/precision=1
            },
            scaled y ticks=false
        ]
        
        \addplot[scatter, only marks, scatter src=explicit symbolic] table[col sep=comma,header=false,x index=6,y index=7, meta index=11]{figures/eval_efficientnet_result.tex};
        \node[coordinate,pin={[pin distance=0.2cm]90:{\scriptsize\color{orange}EYR}}] at (axis cs:114.37373444961958,1.8216538904971054) {};
        \node[coordinate,pin={[pin distance=0.15cm]-90:{\scriptsize\color{blue}SMB}}] at (axis cs:87.6210838927631,3.2627776162884126) {};
        \node[coordinate,pin={[pin distance=0.3cm]0:{\scriptsize\color{teal}Conv\_45}}] at (axis cs:91.46442500592714,4.8139638842950285) {};
    
        \legend{Dominated Points, Non-Dominated Points}
        \end{axis}
    \end{tikzpicture}
    \end{subfigure}\hfill
    \begin{subfigure}[t]{\figrelwidth\linewidth}
    \centering
    \caption{\footnotesize{EfficientNet-B0}}
    \begin{tikzpicture}
        \begin{axis}[
            xlabel={\scriptsize{Top-1 Accuracy [\%]}},
            ylabel={\scriptsize{Throughput[FPS]}},
            width=\figwidth cm,
            height=\figheight cm,
            scatter/classes={%
                dom={mark=x,scale=0.85,draw=black!60},
                nondom={mark=oplus,scale=0.85,draw=red, fill=red},
                sensor={mark=square*, scale=1,draw=orange ,fill=orange},
                edge={mark=square*, scale=1,draw=blue,fill=blue},
                opt={mark=triangle*, scale=2,draw=teal,fill=teal},
                opt2={mark=triangle*, scale=2,draw=olive,fill=olive}
            },
            x label style={at={(axis description cs:0.5,-0.1)}, anchor=north},
            y label style={at={(axis description cs:-.25,0.5)}, anchor=north},
            tick label style={font=\scriptsize},
            yticklabel style={
                /pgf/number format/fixed,
                /pgf/number format/precision=1
            },
             x filter/.code={\pgfmathparse{#1*100}\pgfmathresult},
            scaled y ticks=false
        ]
        
        \addplot[scatter, only marks, scatter src=explicit symbolic] table[col sep=comma,header=false,x index=1,y index=0, meta index=2]{figures/eval_efficientnet_result_acc.tex};
        \node[coordinate,pin={[pin distance=0.25cm]90:{\scriptsize\color{orange}EYR}}] at (axis cs:76.24512,1.8216538904971054) {};
        \node[coordinate,pin={[pin distance=0.25cm]-90:{\scriptsize\color{blue}SMB}}] at (axis cs:75.57373,3.2627776162884126) {};
        \node[coordinate,pin={[pin distance=0.25cm]-90:{\scriptsize\color{teal}Mul\_23}}] at (axis cs:76.80664,4.053987437179409) {};

        \end{axis}
    \end{tikzpicture}
    \end{subfigure}\hfill
    \caption{Selected exploration results for a system consisting of an Eyeriss-like (EYR) accelerator in platform A and a Simba-like (SMB) accelerator in platform B linked via Gigabit Ethernet. The orange and blue squares mark the cases in which the inference is performed either completely on platform A or B, triangles highlight beneficial solutions.}
    \label{fig:eval}
\end{figure*}
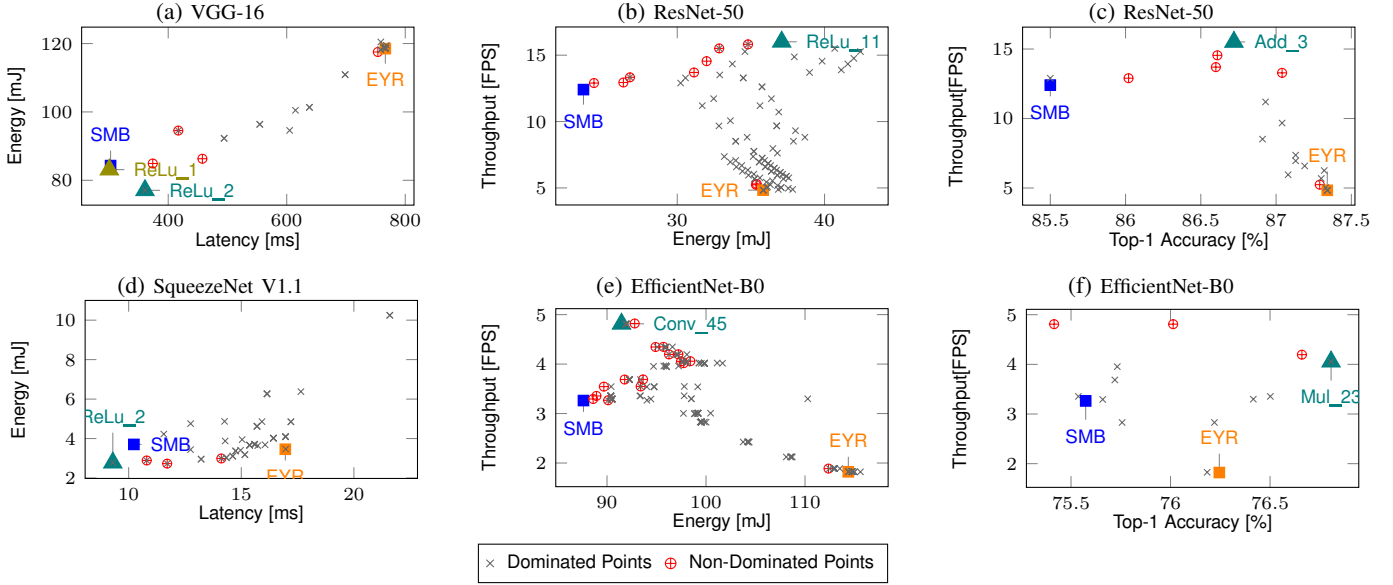

\subsection{Experimental Results}
The simulation time of our proposed framework is dominated by the optional retraining time as expected.
In particular, for models such as EfficientNet-B0, performing retraining for each potential partitioning point takes each about one hour on our system using 32 parallel worker threads for the data loader.
In contrast, graph analysis and hardware evaluation together take approx. 40~min for EfficientNet-B0. 
However, quantization to 16 or 8-bit usually only leads to a slight loss of accuracy, which is why retraining several epochs is often not necessary. 
Nevertheless, we will show below how retraining can affect the partitioning decision for ResNet-50 and EfficientNet-B0.
As shown in \autoref{fig:eval}(a) for VGG-16, partitioning the inference can lead to lower overall energy consumption of the system if the second ReLu is chosen as the partitioning point. 
Furthermore, it is even possible to achieve lower latency and lower energy consumption when partitioning the inference at \textit{ReLu\_1} compared to running the entire \gls{cnn} on SMB. 
Similar results can be obtained for SqueezeNet V1.1, as shown in \autoref{fig:eval}(d).
In this case, energy consumption and latency are significantly reduced by choosing \textit{ReLu\_2} as the partitioning point.
In terms of throughput, as expected, the use of two accelerators allows higher values to be achieved when pipelining is used. 
For ResNet-50, as can be seen in \autoref{fig:eval}(b), the highest possible throughput is achieved with \textit{ReLu\_11}, which requires only slightly more energy than if the entire inference were performed on EYR. 
An increase in throughput of 29\% is achieved. 
A significantly larger increase of 47.5\% can be observed for EfficientNet-B0 at the partitioning point \textit{Conv\_45} (see \autoref{fig:eval}(e)).
This point also offers only marginally higher energy consumption compared to the pure execution of the inference on SMB.
However, the results of the evaluation also show that the throughput can drop significantly if the partitioning point is not chosen carefully. 
This shows the advantages of our approach over AxoNN \cite{axonn} and CNNParted \cite{cnnparted}, which do not explicitly include throughput in their search.
The two accelerators EYR and SMB differ in particular in the bit width, which can have an influence on the accuracy of \glspl{dnn}. 
For this reason, this metric must also be taken into account during evaluation of inference partitioning. 
In this regard, the results for ResNet-50 (\autoref{fig:eval}(c)) and EfficientNet-B0 (\autoref{fig:eval}(f)) show that, as expected, partitioning leads to an improvement in top-1 accuracy compared to running the network entirely on SMB. 
As a guideline, the later the partitioning of the network is performed, i.e. the more layers are executed on EYR, the higher the top-1 accuracy.
However, especially in the case of ResNet-50, the throughput suffers from a later partitioning point as shown in \autoref{fig:eval}(b). 
For this reason, it is necessary at this point to carefully balance the two metrics depending on the use case.

\begin{figure}[t]
    \centering
    \begin{tikzpicture}
    \pgfplotstableread[col sep=semicolon]{./figures/efficientnet_mem.csv}\datatable
        \begin{axis}[
            width=\columnwidth,
            height=3.7cm,
            ymin=0,
            ylabel={\scriptsize{Memory [MB]}},
            xlabel={\scriptsize{Partitioning point}},
            xticklabels from table = {\datatable}{Layer}, 
            xtick=data,
            xticklabel style={rotate=90, font=\tiny},
            enlarge x limits=0.05,
            yticklabel style={
                font=\tiny,
                /pgf/number format/fixed,
                /pgf/number format/precision=1
            },
            y filter/.code={\pgfmathparse{#1/1000000}\pgfmathresult},
            legend style={
                legend columns=1,
                nodes={scale=0.6, transform shape},
                at={(0.6,0.25)},
                anchor=center,
                draw=none,
                fill=none,
                column sep=0.2em
            }
        ]
        \addplot[thick, KITred_70, mark=*, mark options={scale=0.75}] table [x expr=\coordindex, y index=6] {\datatable};
            \addlegendentry{Platform A}
        \addplot[thick, KITblue_100, mark=diamond*, mark options={scale=0.85}] table [x expr=\coordindex, y index=7] {\datatable};
            \addlegendentry{Platform B}
    
        \end{axis}
    \end{tikzpicture} 
    \caption{EfficientNet-B0 results of the analysis of memory resources for a system consisting of two 16-bit platform architectures A and B.}
    \label{fig:eval_mem}
\end{figure}
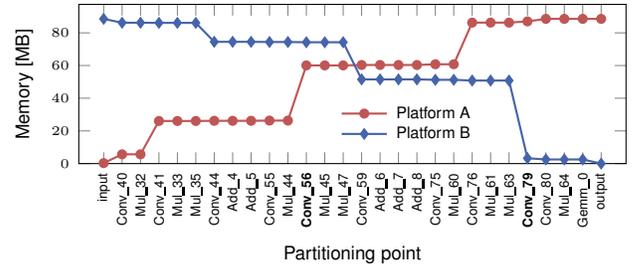

Finally, \autoref{fig:eval_mem} shows the required memory size for executing partitioned \gls{dnn} inference on two 16-bit hardware accelerators A and B.
When compared to other \glspl{dnn} where the overall memory utilization is dominated by the first layers, the memory size required for EfficientNet-B0 increases the later the partitioning in the \gls{dnn} is performed. 
In this case, it is preferable to select a layer before \textit{Conv\_56} or after \textit{Conv\_79} in order to reduce the required system memory and thus  the required die area.
Consequently, the results show the importance of evaluating memory consumption for \gls{dnn} inference partitioning in embedded systems.

\subsection{Increasing Number of Partitioning Points}
In certain areas, such as the automotive sector, the embedded system may have additional computing platforms between the sensor nodes and the central unit, such as the zonal gateway.
These can also be equipped with a \gls{dnn} accelerator.  
As a result, we also evaluate the impact on using more than a single partitioning point.
For our evaluation, we assume a system consisting of two EYR-based platforms in the beginning and two SMB-based platforms at the end of the chain of platforms, connected each via Gigabit Ethernet.
We configure our framework to find pareto-optimal points regarding overall latency, energy consumption and link bandwidth.
The results of our exploration are shown in \autoref{tab:multi_points}.

Especially in small \glspl{dnn}, the use of all available accelerators for inference does not seem to have any practical benefit. 
The reason for this is the high transmission costs, which lead to poorer latency and lower energy consumption. 
Larger \glspl{dnn} such as RegNetX-400MF and EfficientNet-B0, on the other hand, benefit from a system architecture consisting of more platforms, as a significantly higher throughput can be achieved.
These findings demonstrate again that, in contrast to the approach and the experimental results presented by the authors of AxoNN \cite{axonn}, the consideration of other metrics in addition to latency and energy consumption is essential to find a performant and efficient inference partitioning of \glspl{dnn} in distributed embedded systems.

\begin{table}[t]
    \renewcommand{\arraystretch}{1.2}
    \centering
    \scalebox{0.75}{
    \begin{tabular}{r|cccc}
        \textbf{Model} & \textbf{1 Partition} & \textbf{2 Partitions} & \textbf{3 Partitions} & \textbf{4 Partitions}  \\
        \toprule
        SqueezeNet V1.1 & 1 & 5 & 7 & 1 \\
        \midrule
        VGG-16 & 2 & 8 & 8 & 2 \\
        \midrule
        GoogLeNet & 2 & 14 & 8 & 2 \\
        \midrule
        ResNet-50 & 2 & 10 & 10 & 5 \\
        \midrule
        RegNetX-400MF & 2 & 6 & 12 & 13 \\
        \midrule
        EfficientNet-B0 & 2 & 11 & 18 & 19 \\
        \bottomrule
    \end{tabular}}
    \vspace{0.2cm}
    \caption{Number of Partitions for inference that near-optimal schedules include for a system consisting of four accelerators connected via Gigabit Ethernet}
    \vspace{-0.5cm}
    \label{tab:multi_points}
\end{table}

\section{Conclusion}
Efficiently mapping \gls{dnn}-based applications onto distributed embedded systems is a complex task that requires the evaluation of several important performance metrics.
To address this problem, in this work we have presented an automated design space exploration framework for hardware-aware inference partitioning of \glspl{dnn} in distributed embedded systems.
The framework takes into account several system constraints to automatically determine a near-optimal schedule for a given \gls{dnn}.
Our experimental results prove the effectiveness of the proposed framework by revealing multiple beneficial partitioning points besides running inference on a single accelerator for different models.
Furthermore, we were able to show that partitioning the \gls{dnn} inference over more than two accelerators can be useful, especially for large \gls{dnn} architectures.
As a result, our results emphasize the importance of holistic hardware/software code design to enable efficient inference in systems with multiple accelerators.

\section*{Acknowledgment}
This work was funded by the German Federal Ministry of Education and Research (BMBF) under grant number 16ME0817 (CeCaS). The responsibility for the content of this publication lies with the authors.

\bibliographystyle{IEEEtran}
\bibliography{biblio}

\end{document}